\pdfoutput=1
\documentclass[times,twocolumn]{aastex631} % twocolumn, manuscript
\usepackage{graphicx}
\usepackage{mathtools,upgreek} 
\usepackage{natbib}
\usepackage{outlines}
\usepackage{amsmath}
\usepackage{appendix}
\usepackage{tabularx}
\usepackage[version=4]{mhchem} % Molecular species! e.g. \ce{H2O}
\usepackage{comment}
\newcommand{\um}{$\upmu$m}

\received{May 16, 2023}
\revised{July 29, 2023}
\accepted{August 7, 2023}

\submitjournal{The Astrophysical Journal Letters}

\newif\iftwocol
\twocoltrue
\begin{document}

\title{Potential Atmospheric Compositions of TRAPPIST-1~c constrained by JWST/MIRI Observations at 15\um{}}

\correspondingauthor{Andrew P. Lincowski}
\email{alinc@uw.edu}

\author[0000-0003-0429-9487]{Andrew P. Lincowski}
\affiliation{Department of Astronomy and Astrobiology Program, University of Washington, Box 351580, Seattle, Washington 98195, USA}
\affiliation{NASA NExSS Virtual Planetary Laboratory, Box 351580, University of Washington, Seattle, Washington 98195, USA}

\author[0000-0002-1386-1710]{Victoria S. Meadows}
\affiliation{Department of Astronomy and Astrobiology Program, University of Washington, Box 351580, Seattle, Washington 98195, USA}
\affiliation{NASA NExSS Virtual Planetary Laboratory, Box 351580, University of Washington, Seattle, Washington 98195, USA}

\author[0000-0003-0562-6750]{Sebastian Zieba}
\affiliation{Max-Planck-Institut f\"ur Astronomie, K\"onigstuhl 17, D-69117
Heidelberg, Germany}
\affiliation{Leiden Observatory, Leiden University, Niels Bohrweg 2, 2333CA
Leiden, The Netherlands}

\author[0000-0003-0514-1147]{Laura Kreidberg}
\affiliation{Max-Planck-Institut f\"ur Astronomie, K\"onigstuhl 17, D-69117
Heidelberg, Germany}

\author[0000-0002-4404-0456]{Caroline Morley}
\affiliation{{Department of Astronomy, University of Texas at Austin, 2515 Speedway, Austin TX 78712, USA}}

\author[0000-0003-1462-7739]{Michaël Gillon}
\affiliation{Astrobiology Research Unit, University of Liège, Allée du 6 août 19, 4000 Liège, Belgium}

\author[0000-0001-9619-5356]{Franck Selsis}
\affiliation{Laboratoire d'astrophysique de Bordeaux, Univ. Bordeaux, CNRS, B18N, all{\'e}e Geoffroy Saint-Hilaire, 33615 Pessac, France}

\author[0000-0002-0802-9145]{Eric Agol}
\affiliation{Department of Astronomy and Astrobiology Program, University of Washington, Box 351580, Seattle, Washington 98195, USA}
\affiliation{NASA NExSS Virtual Planetary Laboratory, Box 351580, University of Washington, Seattle, Washington 98195, USA}

\author[0000-0001-5657-4503]{Emeline Bolmont}
\affiliation{Observatoire astronomique de l'Universit\'e de Gen\`eve, chemin Pegasi 51, CH-1290 Versoix, Switzerland}
\affiliation{Centre Vie dans l’Univers, Universit\'e de Gen\`eve, Geneva, Switzerland}

\author[0000-0002-7008-6888]{Elsa Ducrot}
\affiliation{Université Paris-Saclay, Université Paris-Cité, CEA, CNRS, AIM}
\affiliation{Paris Region Fellow, Marie Sklodowska-Curie Action}

\author[0000-0003-2215-8485]{Renyu Hu}
\affiliation{Jet Propulsion Laboratory, California Institute of Technology, Pasadena, CA, USA}
\affiliation{Division of Geological and Planetary Sciences, California Institute of Technology, Pasadena, CA, USA}

\author[0000-0002-9076-6901]{Daniel D. B. Koll}
\affiliation{{Department of Atmospheric and Oceanic Sciences, Peking University, Beijing, People's Republic of China}}

\author{Xintong Lyu}
\affiliation{{Department of Atmospheric and Oceanic Sciences, Peking University, Beijing, People's Republic of China}}

\author[0000-0002-8119-3355]{Avi Mandell}
\affiliation{NASA Goddard Space Flight Center, 8800 Greenbelt Rd, Greenbelt, MD, USA}
\affiliation{Sellers Exoplanet Environments Collaboration, NASA Goddard}

\author[0000-0003-4471-1042]{Gabrielle Suissa}
\affiliation{Department of Astronomy and Astrobiology Program, University of Washington, Box 351580, Seattle, Washington 98195, USA}
\affiliation{NASA NExSS Virtual Planetary Laboratory, Box 351580, University of Washington, Seattle, Washington 98195, USA}

\author[0000-0003-2171-5083]{Patrick Tamburo}
\affiliation{Department of Astronomy \& The Institute for Astrophysical Research, Boston University, 725 Commonwealth Ave., Boston, MA 02215, USA}

\shorttitle{TRAPPIST-1 c Secondary Eclipses Analysis}
\shortauthors{Lincowski et al.}

\begin{abstract}
The first JWST observations of TRAPPIST-1~c showed a secondary eclipse depth of 421$\pm94$~ppm at 15~\um{}, {which is consistent with a bare rock surface or a thin, \ce{O2}-dominated, low \ce{CO2} atmosphere \citep{Zieba:2023}.} Here, we further explore potential atmospheres for TRAPPIST-1~c by comparing the observed secondary eclipse depth to synthetic spectra of a broader range of plausible environments.{ To self-consistently incorporate the impact of photochemistry and atmospheric composition on atmospheric thermal structure and predicted eclipse depth, we use a two-column climate model coupled to a photochemical model, and simulate \ce{O2}-dominated, Venus-like, and steam atmospheres.} We find that a broader suite of plausible atmospheric compositions are also consistent with the data.  For lower pressure atmospheres (0.1 bar), our \ce{O2}-\ce{CO2} atmospheres produce eclipse depths within 1$\sigma$ of the data, consistent with the modeling results of \citet{Zieba:2023}. {However, for higher-pressure atmospheres, }our models produce different temperature-pressure profiles and are less pessimistic, with 1--10~bar \ce{O2}, 100~ppm \ce{CO2} models within 2.0--2.2$\sigma$ of the measured secondary eclipse depth, and up to 0.5\% \ce{CO2} within 2.9$\sigma$.  {Venus-like atmospheres are still unlikely.}  For thin \ce{O2} atmospheres of 0.1~bar with a low abundance of \ce{CO2} ($\sim$100~ppm), up to 10\% water vapor can be present and still provide an eclipse depth within 1$\sigma$ of the data.  {We compared the TRAPPIST-1 c data to modeled steam atmospheres of $\leq$ 3~bar, which are 1.7--1.8$\sigma$ from the data and not conclusively ruled out. More data will be required to discriminate between possible atmospheres, or to more definitively support the bare rock hypothesis.}

\end{abstract}

\section{Introduction}

\textit{James Webb Space Telescope} (JWST) observations of the TRAPPIST-1 planetary system \citep{Gillon:2017,Luger:2017a} are now providing the first opportunity to search for and probe the atmospheres of truly Earth-sized planets outside the Solar System
\citep[e.g.][]{Greene:2023,Zieba:2023}.  The targets for these proposals included all seven planets of the TRAPPIST-1 system, and the orbital distances of these planets span and extend beyond the limits of the habitable zone. This system of planets is therefore ideal for understanding terrestrial planetary evolution and habitability, and for initiating the search for life on exoplanets \citep{Lincowski:2018,Lustig:2019,Meadows:2020}.

Recently, JWST/MIRI \citep{Wright:2023} {secondary eclipse measurements were }used to provide the first observational constraints on {whether or not the two innermost planets, TRAPPIST-1 b and c, had atmospheres \citep{Greene:2023,Ih:2023,Zieba:2023}}.  {These studies complement previous efforts to use secondary eclipse measurements to probe atmospheric composition and thickness on hot rocky exoplanets \citep{Kreidberg:2019,Whittaker:2022,Crossfield:2022}}. {For TRAPPIST-1~b, } JWST Program GTO1177 obtained five secondary eclipse observations, with a measured depth of 861$\pm$99~ppm in the MIRI F1500W filter \citep{Greene:2023}. For TRAPPIST-1~c, which receives a similar insolation to Venus in our planetary system, JWST program GO2304 obtained four secondary eclipse observations, with a measured eclipse depth of 421$\pm94$~ppm \citep {Zieba:2023}.  The F1500W filter was selected for atmosphere detection for both planets, as it is is sensitive to absorption from the 15~\um{} \ce{CO2} band, which is prevalent in the spectrum of the atmosphere-bearing terrestrial planets of our Solar System (Venus, Earth, and Mars). \citet{Greene:2023} found that the 15~\um{} secondary eclipse depth for TRAPPIST-1~b was most consistent with a dark airless rock in thermal equilibrium on the day-side, with little to no heat redistribution from the day to night side. They conclusively ruled out both a Venus-like atmosphere and an \ce{O2}-dominated atmosphere with 0.5 bars of \ce{CO2} to greater than 6$\sigma$. {Follow-on interpretation of the TRAPPIST-1 b observations using a self-consistent radiative-convective equilibrium model suggested that plausible atmospheres with at least 100~ppm of \ce{CO2} were ruled out at 3$\sigma$ for pressures greater than 0.3~bar \citep{Ih:2023}, and that thicker atmospheres were only possible in the unlikely event that the atmosphere lacks any strong MIR absorbers.}  For TRAPPIST-1~c, the \citet{Zieba:2023} 15~\um{} secondary eclipse depth of {421$\pm94$~ppm} lacks the precision to conclusively determine whether the planet has an atmosphere or is a bare rock. \citet{Zieba:2023} found that ultramafic rock was consistent within 1$\sigma$ of the secondary eclipse depth, as were several \ce{O2} atmospheres with low abundances of \ce{CO2}, such as 0.1~bar \ce{O2} with 100~ppm \ce{CO2}. They showed Venus-like and thick \ce{O2}-\ce{CO2} atmospheres ($\ge$10~bar) were unlikely (at $\geq$ $2.6\sigma$), as well as 1~bar atmospheres with \ce{CO2} abundances $\geq$ 1000~ppm, which were ruled out to $\geq3\sigma$. 

However, the \citet{Zieba:2023} initial analyses did not include other plausible environments for TRAPPIST-1 c that have yet to be compared to the constraints provided by the secondary eclipse data. TRAPPIST-1~c has a lower density than the Earth, suggesting a currently volatile-rich or iron-poor interior \citep{Grimm:2018,Agol:2021}, and is expected to have been subjected to high levels of radiation early in its history, which could have driven atmospheric escape and water loss \citep{Bolmont:2017,Dong:2018,Lincowski:2018,Wordsworth:2018,Zieba:2023}. If water loss is ongoing, then the planet could be in a constant runaway greenhouse state, with a water-dominated steam atmosphere \citep{Turbet:2020}. If water loss was extensive and nearly complete, a Venus-like or oxygen-dominated atmosphere could exist \citep{Luger:2015b,Meadows:2018a,Lincowski:2018,Wordsworth:2018}. Moreover, while \citet{Zieba:2023} considered a grid of \ce{O2}-\ce{CO2} atmospheres in their initial assessment, they used only those two gases and simplified temperature profiles to predict their eclipse depths. The prescribed temperature profiles, consisting of tropospheric adiabats with isotherms above 0.1~bar (or the skin temperature), were modeled in thermal equilibrium with insolation, as in \citet{Morley:2017}, with an adjustment for day-night heat redistribution \citep{Koll:2019}. However, more realistic atmospheres would include a cocktail of outgassed constituents, as well as photochemical byproducts. As discussed by \citet{Lincowski:2018}, these additional species could alter the atmospheric temperature structure (e.g. by forming a stratospheric temperature inversion) and strongly impact the predicted secondary eclipse depths. This may then impact our  confidence in whether a given environment is consistent with the data.  

Here we extend the analysis of \citet{Zieba:2023} with a broader and more self-consistent assessment of plausible atmospheres using a 1.5D (two-column, day-night with heat transport) climate model coupled to a photochemical model.  This analysis improves on the initial assay of \citet{Zieba:2023}, by taking into account photochemistry and atmospheric composition when determining the atmospheric temperature structure and eclipse depths. We calculate and present day-side temperatures, brightness temperature spectra, and associated secondary eclipse depths for Venus-like (\ce{CO2}-dominated), water-dominated (steam), and \ce{O2}-dominated environments. We also include three examples to compare with the grid of two-component \ce{O2}-\ce{CO2} atmospheres considered in  \citet{Zieba:2023}. We discuss our results, and suggest future observations that may help determine or exclude potential atmospheres for TRAPPIST-1~c. 

\clearpage

\section{Methods}

Here we use the two-column, day-night capabilities of our versatile 1D radiative-convective-equilibrium, coupled climate-photochemical model for terrestrial planets, VPL Climate \citep{Robinson:2018,Lincowski:2018}, to compute the atmospheric states and corresponding secondary eclipse depths for our simulations. Notably, \citet{Robinson:2018} conducted a full climate validation for Venus with VPL Climate, while \citet{Lincowski:2018} successfully modeled the middle atmosphere of Venus using the updated photochemical model. 

VPL Climate uses the Spectral Mapping Atmospheric Radiative Transfer code (SMART) for radiative transfer. SMART is a spectrum-resolving, multistream, multi-scattering model developed by D. Crisp \citep{Meadows:1996,Crisp:1997} that uses the Discrete Ordinate Radiative Transfer code \citep[DISORT,][]{Stamnes:1988,Stamnes:2000} to compute the radiation field. To compute the solar heating rates on the day side, we conduct four heating rate calculations at distinct angles spanning solar zenith angles of 21--86~degrees, and integrate these using Legendre-Gauss quadrature. For recent and extensive descriptions of the full capabilities of SMART, see \citet{Meadows:2018a}, \citet{Robinson:2018}, and \citet{Lincowski:2018}.
VPL Climate incorporates latent heating and cooling rates due to condensable gases (here either water or sulfuric acid). 
Vertical transport is specified through a mixing length parameterization. Advective mixing between the day and night side is calculated layer-by-layer, based on a two-column closure of the 3D primitive equations for global transport. The vertical and horizontal transport, along with associated parameter choices, are explained in detail by \citet{Lincowski:PhD} and discussed briefly in Appendix~\ref{app:twocol}, which also provides some model parameters and some validation comparisons with 3D GCM results.

The photochemical-kinetics model coupled to our 1D climate model is described in \citet{Lincowski:2018}. The photochemical model computes photolysis and kinetic reactions using 200 plane-parallel layers, with diffusion and eddy transport. The model includes diffusion-limited top-of-atmosphere escape for hydrogen. The model includes condensation of water and sulfuric acid, and includes some aqueous phase chemistry and rainout. Aerosols for both models are calculated as described in \citet{Meadows:2018a}, \citet{Lincowski:2018}, and \citet{Meadows:2020}.

Absorption lines associated with visible to mid-infrared (MIR) transitions are calculated using the line-by-line model, LBLABC \citep{Meadows:1996}, using the HITRAN2016 \citep{HITRAN:2016}, HITEMP \citep{HITEMP:2010}, or Ames \citep{Huang:2017} line databases. For a list of collision-induced absorption and UV cross sections, see \citet{Lincowski:2018}.

\subsection{Model Inputs}

\begin{table*}[thb!]
    \centering
    \caption{\textbf{\textsc{Modeled Planetary States and Their Environmental Parameters}}.}
    \label{table:experiments}
    {\footnotesize\selectfont
    \begin{tabularx}{\textwidth}{lcc>{\centering\arraybackslash}c>{\centering\arraybackslash}p{2.4cm}>{\centering\arraybackslash}p{1.45cm}>{\centering\arraybackslash}p{1cm}>{\centering\arraybackslash}p{1.75cm}>{\centering\arraybackslash}p{1.75cm}}
    \hline \hline
    \textbf{Environment}  & \textbf{\ce{CO2}} & \textbf{Surf. Pres.} & \textbf{Aerosols}  & \textbf{Day / Global / Night Surf. Temp. [K]}   & \textbf{Ecl. Depth$^a$ [ppm]} & \textbf{Dev.$^b$ [$\sigma$]} & \textbf{Day / Night Brightness Temp.$^a$ [K]} & {Day / Night OLR$^d$ [W~m$^{-2}$]} \\
    \hline
        Venus-like          & 96.5\% &    0.1 bar    &   None           & 426 / 365 / 244 & {136} & 3.0 & 270 {/ 233} & {1139 / 276} \\ 
        Venus-like          & 96.5\% &  1 bar      &   None           & 485 / 444 / 383 & {128} & {3.1} & 265 {/ 227} & {868 / 574} \\ 
        Venus-like          & 96.5\% &    10 bar     &   None           & 643 / 623 / 601 & {131} & {3.1} & 267 {/ 225} & {754 / 707} \\ 
        Venus-like          & 96.5\% &   10 bar     &   \ce{H2SO4}     & 602 / 601 / 570 & {173} & 2.6 & 288 {/ 244} & {642 / 514} \\
        Steam               & 1000~ppm &  0.1 bar    &   night cirrus   & 547 / 506 / 433 & {256} & {1.8} & 323 {/ 285} & {1007 / 529} \\
        Steam               & 1000~ppm &   1 bar      &   night cirrus   & 645 / 628 / 610 & {256} & {1.8} & 323 {/ 282} & {941 / 612} \\
        Steam               & 1000~ppm &   3 bar      &   night cirrus   & 705 / 678 / 693 & {260} & {1.7} & {325} {/ 285} & {962 / 593} \\
        \ce{O2}-\ce{H2O}    & 100~ppm &   0.1 bar    &   None           & 448 / 384 / 267 & 331 & 1.0 & 350 {/ 291} & {1111 / 351} \\
        {\ce{O2 pure}}             & --- &   {0.1 bar}    &   {None}           & {395 / 335 / 156} & {444} & {0.2} & {392 / 165}  & {1341 / 45} \\
        \ce{O2}             & 100~ppm &   0.1 bar    &   None           & 419 / 358 / 235 & {369} & {0.6} & 365 {/ 264}  & {1180 / 259} \\
        \ce{O2}             & 100~ppm &   1 bar      &   None           & {453} / {415} / {356} & {238} & 2.0 & {316 / 304} & {886 / 599} \\
        \ce{O2}             & 100~ppm &   10 bar     &   None           & 493 / {478} / 457 & {211} & 2.2 & 305 {/ 295} & {785 / 699} \\
        \ce{O2}             & 500~ppm &   10 bar     &   None           & {517} / {495} / {479} & {184} & {2.5} & {293 / 281} & {803 / 712} \\
        \ce{O2}             & 0.5\% &    10 bar     &   None           & {542} / 527 / {511} & {152} & 2.9 & {278 / 261}  & {806 / 716} \\ \hline
        \ce{O2}$^c$         & 100~ppm &   0.1 bar    &   None           & 470 / --- / --- & 481 & 0.6 & 399 / --- & { / } \\
        \ce{O2}$^c$         & 100~ppm &    1 bar      &   None           & 461 / --- / --- & 200 & 2.4 & 296 / --- & { / } \\
        \ce{O2}$^c$         & 100~ppm &   10 bar     &   None           & 461 / --- / --- & 48  & 4.0 & 206 / --- & { / } \\
        \hline 
    \end{tabularx}
    \flushleft
    $^\text{a}$ Day-side, integrated over JWST/MIRI F1500W band. \\
    $^\text{b}$ Deviation from the mean measurement of \citet{Zieba:2023} using their measurement error. \\
    $^\text{c}$ Model and results from \citet{Zieba:2023}. {Note the surface albedo was 0.1.} \\
    $^\text{d}$ Outgoing longwave radiation. \\
    }
\end{table*}

The model atmospheres (listed in Table~\ref{table:experiments}) represent a selection of plausible atmospheres for TRAPPIST-1~c, given likely stellar and atmospheric evolution \citep{Lincowski:2018}. These include steam atmospheres, post-ocean-loss \ce{O2}-dominated atmospheres, and Venus-like atmospheres.  Note that the majority of these atmospheres are assumed to contain at least trace amounts of 
\ce{CO2}. {Given that these planets have measured densities comparable to Solar System terrestrial planets (albeit slightly lower, which suggests the possibility of significantly higher volatile contents \citealt{Agol:2021}), we have assumed that these planets outgas a suite of gases, including \ce{CO2}. \ce{CO2} outgassing is likely common for terrestrial planets in the magma ocean stage, over a broad range of redox states \citep{Gaillard2022}. \ce{CO2} is also likely to continue to be a principal component of magmatic outgassing (along with water vapor) for overlying atmospheric pressures in excess of 0.1~bar, although \ce{SO2} outgassing also becomes progressively more important at lower pressures \citep{Gaillard2014}.}  

The atmospheres are plane-parallel and contain 32--64 pressure levels.  The top of each model atmosphere extends to 0.01~Pa for inclusion of appropriate photochemistry. We use the {nominal planetary and stellar parameters} for TRAPPIST-1~c {from \citet{Agol:2021}, derived also from \citet{Mann:2019} and \citet{Ducrot:2020}}.

Our consideration of steam atmospheres is based on the likelihood that the TRAPPIST-1 planets formed exterior to the snow-line and migrated inwards to their current positions and so may currently be volatile rich \citep{Luger:2017a,Grimm:2018,Agol:2021}. 
The steam atmospheres are assumed to have a water mixing ratio of 1 at the surface, but to also contain other constituents from ongoing interior outgassing. The outgassing is assumed to maintain the atmosphere against loss processes, which may be plausible if outgassing flux replenishment rates exceed loss rates, as has been argued for other M~dwarf planets \citep{Garcia:2017}. Assuming a volatile-rich interior, we use the following {plausible} outgassing fluxes that are in a ratio more reduced than Earth:  \ce{H2} ($3\times10^{10}$~molecules~s$^{-1}$~cm$^{-2}$), CO ($2\times10^{8}$~molecules~s$^{-1}$~cm$^{-2}$), and \ce{CH4} ($6.8\times10^{8}$~molecules~s$^{-1}$~cm$^{-2}$, \citealt{Guzman:2013}). These atmospheres have 1000~ppm (0.1\%) \ce{CO2}.

Our oxygen-dominated atmospheres include a range of pressures and outgassing fluxes. {Here we use outgassing for all except the pure \ce{O2} case, to complement the results of \citet{Zieba:2023}, who assumed dry atmospheres}. We employ a 0.1~bar, 100~ppm \ce{CO2} atmosphere for comparison with the best atmospheres from the \ce{O2}-\ce{CO2} grid presented by \citet{Zieba:2023}, but we include the same volcanic fluxes as in our steam atmospheres, with an \ce{H2O} flux of $1.68\times10^{11}$~molecules~s$^{-1}$~cm$^{-2}$, as in \citet{Lincowski:2018}.  To test the climatic and spectral impact of additional water vapor in these atmospheres, we also modeled a 0.1~bar \ce{O2}, 100~ppm \ce{CO2} atmosphere that contains 10\% water vapor, labeled as \ce{O2}-\ce{H2O} in Table \ref{table:experiments}. To represent the other end of the \citet{Zieba:2023} \ce{O2}-\ce{CO2} grid, we include 1 and 10~bar \ce{O2} atmospheres with 100~ppm \ce{CO2}, and the same volcanic outgassing. We also model two other 10~bar \ce{O2} atmospheres, with 0.5\% \ce{CO2} \citep{Lincowski:2018}, and 500~ppm \ce{CO2}. {Lastly, we include a pure \ce{O2} atmosphere to have an example with no \ce{CO2} and no other outgassed species. This atmosphere still involves a minimal level of hydrogen, and can generate ozone photochemically.}

We model a range of Venus-like atmospheres derived  from \citet{Lincowski:2018}. We include 0.1, 1, and 10~bar clear-sky atmospheres, and a 10~bar hazy version. All of these use the same surface boundary conditions as described in \citet{Lincowski:2018} and include \ce{CO2}, \ce{N2}, \ce{H2O}, \ce{H2}, NO, \ce{SO2}, OCS, and HCl. Photochemistry modifies the profiles of these gases and their byproducts, including CO and \ce{H2SO4}. These atmospheres were included in the analysis of \citet{Zieba:2023}.

{For the surface of all modeled environments, we use wavelength-dependent reflectance for basalt}
from the U.S.G.S. spectral library \citep{Clark:2007}\footnote{\label{foot:usgs}\url{https://doi.org/10.3133/ds231}}.

Because we require stellar UV as input to the photochemical model, the stellar SED used in this work was derived from \citet{Peacock:2019} and was described in \citet{Meadows:2020}. Critically, this {stellar spectrum includes UV wavelengths} that incorporate stellar UV activity, and that are calibrated to available UV photometric constraints for TRAPPIST-1 \citep{Peacock:2019}. For climate and spectral modeling, we retain this spectrum at its native resolution. In the photochemical model, the spectrum is binned to 100~cm$^{-1}$.  

\begin{figure*}
\centering
\includegraphics[width=\textwidth]{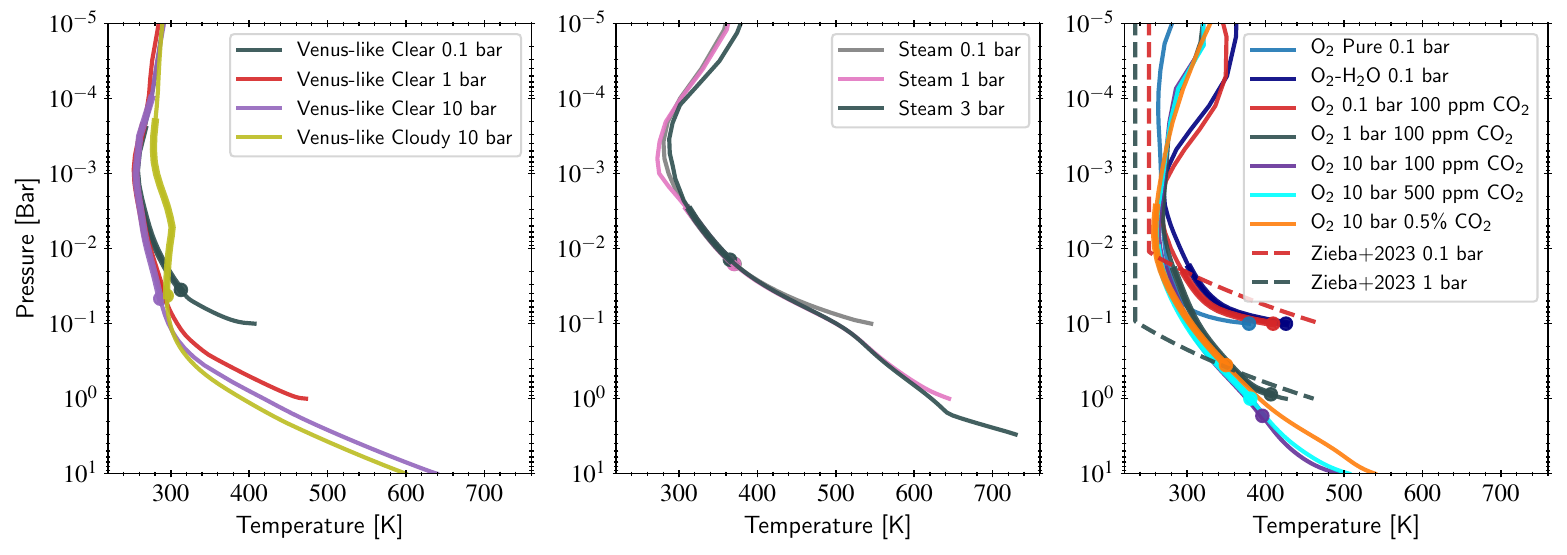}
\caption{Day-side hemisphere temperature structures for all modeled atmospheres: Venus-like (left panel), steam atmospheres (middle panel) and \ce{O2}-\ce{CO2} atmospheres (right panel). {For our modeled atmospheres, we have used a thicker line to show the layers over which the 15~\um{} band reached optical depth of 1 (indicating the effective emission layer), for each model atmosphere. The thick dots indicate the lowest layer probed in the 15~\um{} band.} In the right panel, we also included two example temperature-pressure profiles from \citet{Zieba:2023} for 0.1 and 1 bar \ce{O2}-dominated atmospheres, each with 100~ppm \ce{CO2}. A wide variety of temperature profiles are possible for TRAPPIST-1~c under different atmospheric compositions. Although surface temperatures differ widely, note the similarities in {emission temperatures/pressures probed} for similar atmospheric compositions. }
\label{fig:atms}
\end{figure*}

\subsection{Thermal Emission Spectra and Secondary Eclipse Depths}

We used SMART to produce thermal emission spectra of the day-side hemispheres of our modeled atmospheres and used these to calculate brightness temperature spectra. 
The JWST data of \citet{Zieba:2023} indicated that the brightness temperature of TRAPPIST-1 in the 15~\um{} filter band did not match the \citet{Peacock:2019} stellar model, and so we adjusted it to the measured brightness temperature of 1867~K \citep{Zieba:2023} in this band{ (the factor using our stellar model was 1.28)}. Secondary eclipse depths are calculated by dividing the band-integrated planetary {photon} flux by the band-integrated stellar {photon} flux; this is the fractional amount of flux that disappears from the observer's line of sight when the planet is occulted by the star.

\clearpage 

\section{Results}

To further explore how the 15~\um{} secondary eclipse measurement by \citet{Zieba:2023} constrains the possible presence and nature of an atmosphere for TRAPPIST-1~c, we produced day-night atmospheres for a variety of planetary environments (Table~\ref{table:experiments}). These environments consist of Venus-like (\ce{CO2}-dominated), steam (\ce{H2O}-dominated),  and post-ocean-loss/oxidized \ce{O2}-dominated atmospheres \citep{Luger:2015b,Lincowski:2018}.

The day-side temperature structures are shown in Figure~\ref{fig:atms}.  From the day-side equilibrium states, we produced emission spectra presented as brightness temperature. We used these spectra, convolved with the filter response of the JWST/MIRI F1500W 15~\um{} band, to compare our modeled secondary eclipse depths with the measurement of \citet{Zieba:2023}. Our secondary eclipse predictions are listed in Table~\ref{table:experiments} with the environments modeled.

{Although it is not readily apparent in the JWST 15~\um{} band due to \ce{CO2} absorption there, our models generally follow the intuition that day-night heat transport increases as surface pressure is increased. This is demonstrated in the drop in day-side outgoing longwave radiation (OLR), and the increase in night-side OLR for the Venus-like (276--707~W~m$^{-2}$) and \ce{O2}-dominated atmospheres (e.g. with 100~ppm \ce{CO2}, 259--699~W~m$^{-2}$). The steam atmospheres, even at 0.1~bar, are optically thick and change little with added surface pressure (529--612~W~m$^{-2}$).}

\begin{figure*}
\centering
\includegraphics[width=\textwidth]{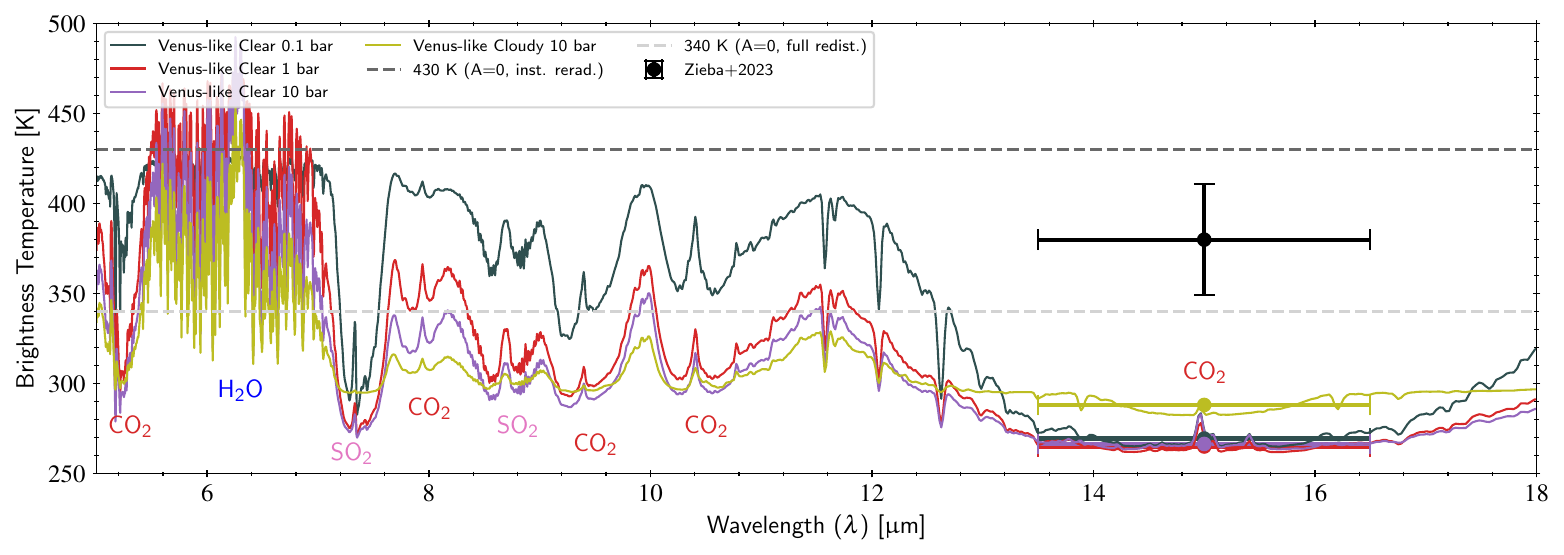}
\includegraphics[width=\textwidth]{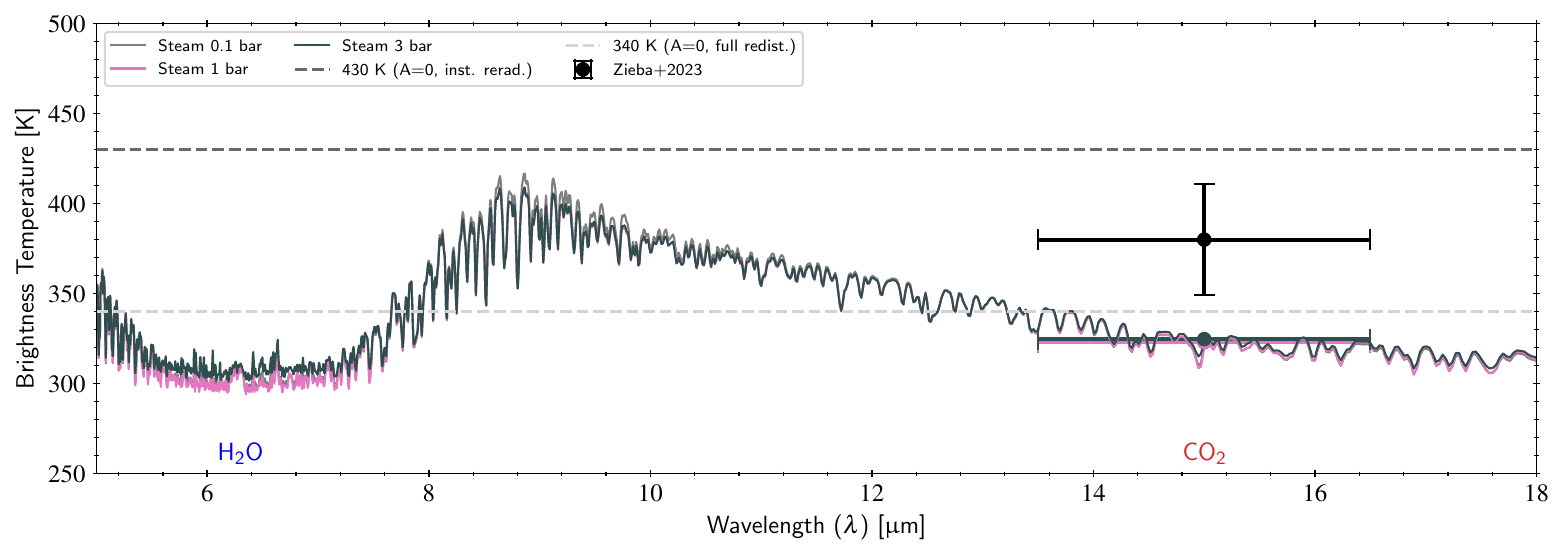}
\includegraphics[width=\textwidth]{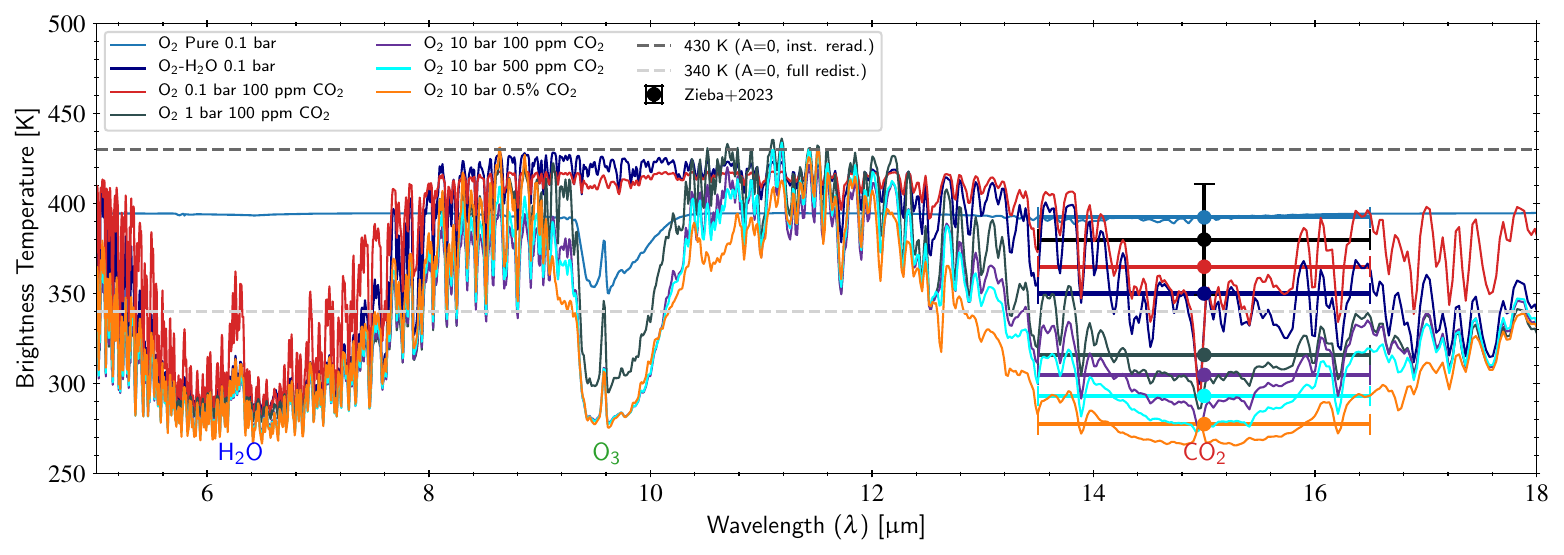}
\caption{Brightness temperature spectra for the dayside hemisphere of all modeled environments, with points corresponding to the model spectra convolved to the F1500W filter band over the band's wavelength extent (horizontal error bars show the FWHM of the filter band). We also plot lines for 340~K (blackbody, no atmosphere, full heat redistribution), and 430~K (blackbody, no atmosphere, no heat redistribution), along with the data point measured by \citet{Zieba:2023}, which has a brightness temperature equivalent to $380\pm31$ K. \textit{Top:} Venus-like modeled environments. \textit{{Middle}:} steam environments. \textit{Bottom:} \ce{O2}-dominated environments. The Venuses are between 2.6--3.1$\sigma$ from the measured eclipse depth and the steam atmospheres are within {1.7--1.8$\sigma$} of the measured eclipse depth. The clear-sky Venuses all exhibit similar \ce{CO2} bands between 0.1--10 bar. The steam environments spanning 0.1--3~bar are nearly identical across the MIR spectrum. The \ce{O2}-dominated environments, with varying amounts of \ce{CO2}, exhibit the largest range in their spectra, and those of $\geq1$~bar also exhibit strong ozone features. For reference, the F1500W brightness temperature values from \citet{Zieba:2023} for their best-fit desiccated \ce{O2}-\ce{CO2} atmosphere is 395~K, and their best-fit bare rock surface is 420~K.}
\label{fig:eclipse_venus}
\end{figure*}

\subsection{Venus-like Atmospheres}

In Figure~\ref{fig:eclipse_venus} (upper left panel), we compare the secondary eclipse depths for our Venus models, consisting of 0.1, 1, and 10-bar clear-sky Venus-like atmospheres. Initial results for eclipse depths from these atmospheres were presented in \citet{Zieba:2023}, and the spectra are provided here.  In this work, Venus-like includes the important Venus trace gases \ce{H2O}, \ce{SO2}, CO, \ce{OCS}, \ce{H2S}, NO, and \ce{HCl}, in photochemical equilibrium with the modeled TRAPPIST-1 UV spectrum. 

The Venus-like planets had very similar 15~\um{} eclipse depths, with the cloudy Venus exhibiting the largest eclipse depth. The clear-sky atmospheres spanning 0.1--10~bar had eclipse depths between 134--143~ppm, and the cloudy Venus, which included sulfuric acid haze aerosols, had an eclipse depth of 181~ppm, which was still 2.6$\sigma$ from the measurement. These results were also noted in \citet{Zieba:2023}. The nearly identical 15~\um{} secondary eclipse depths for the clear-sky Venuses are likely due to similar minimum day-side temperatures, at around 260~K, and similar lower atmosphere profiles, with the 15~\um{} emission originating approximately at these temperature minima in the atmospheric column.  In contrast, although clouds cool the planet overall, the cloudy Venus produces a larger eclipse depth, implying hotter emission temperatures.  This is because the majority of the emission in the core of the 15~\um{} band originates in and above the cloud deck, where sulfuric acid aerosols absorb NIR radiation and locally warm the atmosphere, producing a weak temperature inversion in the 1--50~mbar (38--58~km) region (Figure~\ref{fig:atms}, left panel). Given the insensitivity to the lower atmosphere for these cases, and the extensive computational time required for more dense atmospheres, we did not model a more Venus-like 93~bar atmosphere. 

The Venus-like atmospheres also have \ce{CO2} absorption at 9.4 and 10.4~\um{}, with additional absorption bands shortward due to \ce{H2O}, \ce{CO2}, and \ce{SO2}.

\subsection{Steam Atmospheres}

The modeled steam environments consisted of 0.1, 1, and 3~bar atmospheres with near 100\% mixing ratio of water at the surface. In all cases, they were hot enough that the relative humidity {throughout most of the atmospheric layers} was well below 100\%. We included 0.1\%~\ce{CO2}, along with outgassing by \ce{H2}, \ce{CH4}, and CO, based on the assumption that interior  outgassing would be the likely source of a continually maintained steam atmosphere, given the high escape rates expected for the TRAPPIST-1 planets \citep{Bourrier:2017loss,Bolmont:2017,Dong:2018,Lincowski:2018,Wordsworth:2018,Krissansen:2022,Zieba:2023}. 

All three modeled steam atmospheres exhibited stratospheric condensation of water to form clouds on the night side, but not the day side {(interestingly enough, a similar cloud dichotomy between day and night sides has been identified for hot steam atmospheres for early Earth or Venus, with a full 3D Global Climate Model, see \citealt{Turbet:2021})}. Therefore, the secondary eclipse spectra here are all clear-sky. {Our climate model does not include a microphysical cloud model; to account for clouds, we have specified thin cirrus (water-ice) clouds in the layers of condensation, which in these atmospheres is approximately 10--200~Pa. The cirrus clouds warm the atmosphere due to ice band absorption of thermal outgoing radiation, which causes them to evaporate; we maintain thin ($\tau=0.05-0.15$) clouds here to balance condensation and localized warming.} Although {condensation (and thereby cloud formation) occurred} higher in the atmosphere, the average night-side surface temperature was still well above the freezing point of water for these three atmospheres (433--743~K){, and even modestly thin cirrus clouds in layers of condensation serve to warm the atmosphere and prevent further condensation}. The relatively high atmospheric pressures and incident stellar radiation also make night-side freezing of the atmosphere onto the surface unlikely \citep{Wordsworth:2015}. 

Similar to the Venus-like atmospheres, the steam atmospheres of a variety of pressures (0.1--3~bar) resulted in very similar eclipse depths, {256--260~ppm} (see Figure~\ref{fig:eclipse_venus},  upper right panel). These depths were also significantly higher than those of the Venus-like atmospheres, {correlating with} hotter emission temperatures, and they deviate from the measured data by only {1.7--1.8$\sigma$}.  The steam atmospheres' nearly identical secondary eclipse spectra across the MIR are the product of weak water vapor opacity that increases with wavelength in the MIR, and temperature structures that are nearly identical across the radiative emission altitudes of the respective atmospheres. 
Note in these cases that the 0.1\% \ce{CO2} had minimal impact on the 15~\um{} band.  This is due to the thermal emission being at {3--20~mbar}, {the opacity of \ce{H2O} exceeds that of \ce{CO2}, and} \ce{CO2} absorption from the column above this pressure is negligible. Aside from the ubiquitous 6.3~\um{} water band, these atmospheres have no prominent absorption features that could distinguish them (c.f. Figures~\ref{fig:eclipse_venus} and \ref{fig:eclipse_selected}).

{Although the 15~\um{} depth is similar to the lower pressure steam atmospheres, the 3~bar atmosphere is thick enough to be in a transitional runaway greenhouse state \citep[see also][]{Nakajima:1992,Kopparapu:2013,Turbet:2019}. At the surface temperatures reported here, the 3~bar atmosphere is still heating in the lower atmosphere. Throughout the typical MIR range (see Figure~\ref{fig:eclipse_venus}, center panel), the steam atmospheres have nearly identical spectra. This is due to being optically thick here, and as the atmosphere heats, stability is achieved when the thermal flux is emitted through atmospheric windows into the NIR. This stability was achieved in the 0.1 and 1~bar atmospheres, but not 3~bar. } 

\subsection{\ce{O2}-dominated Atmospheres}

The oxygen-dominated atmospheres vary considerably in secondary eclipse depths (see Figure~\ref{fig:eclipse_venus}, lower panel, and Table \ref{table:experiments}), spanning {152--444~ppm and are 0.2--2.9$\sigma$ from the observation. The pure \ce{O2} atmosphere produces a 15~\um{} eclipse depth that is higher than the data.  All other models produce eclipse depths with lower values than the data.}   Unlike \ce{CO2} and \ce{H2O}, oxygen has minimal molecular absorption bands,and has no absorption in the 15~\um{} filter band.
Although the outgassed \ce{H2O} accumulates to $\sim$0.5--2\% in our models, the absorption in the 15~\um{} filter band is still dominated by trace \ce{CO2}---and the strength of the absorption is a function of atmospheric \ce{CO2} abundance, total atmospheric pressure, and the atmospheric temperature structure.  Since it is also the strongest greenhouse gas in these atmospheres, even trace amounts of \ce{CO2} have a strong influence on the atmospheric temperature structure, which directly affects the filter band emission temperature.  We note also that the predicted hemispherically-averaged nighttime temperature for these atmospheres (Table \ref{table:experiments}) are well above the condensation point of \ce{O2} (90~K at 1~bar) and \ce{CO2} (194~K at 1~bar), such that atmospheric collapse is unlikely to occur.  However, the two 0.1~bar atmospheres have night time temperatures (235~K, 267~K) below the freezing point of water, which may result in \ce{H2O} removal from the atmosphere over time.

Our model suite included climate-photochemical models with interior outgassing for three \ce{O2}-dominated atmospheres specifically for comparison with the atmospheric models considered by \citet{Zieba:2023}, in addition to 10~bar  \ce{O2} atmospheres with higher \ce{CO2}. Three comparison atmospheres all have 100~ppm \ce{CO2}, with surface pressures of 0.1, 1, and 10~bar. These atmospheres produce eclipse depths spanning {369~ppm} (0.1~bar) to {211~ppm} (10~bar), with corresponding {0.6--2.2$\sigma$} fits to the measurement.  For comparison, the \citet{Zieba:2023} fits to the measurement for the pure \ce{CO2}-\ce{O2} atmospheres spanned 0.6--4$\sigma$. Our 10~bar atmosphere with 500~ppm of \ce{CO2} predicts a {184~ppm} eclipse depth (2.6$\sigma$ from the data) and with 0.5\% \ce{CO2}, a 152~ppm eclipse depth that is 2.9$\sigma$ from the data. {Also, as noted above, our pure \ce{O2}, no \ce{CO2} atmospheres produce eclipse depth values that are higher than the data, and \citet{Zieba:2023} saw similar behavior for their low pressure atmospheres ($\leq$0.1~bar) \ce{O2} atmospheres with $\leq$ 100~ppm  \ce{CO2}.}

Other molecular bands were present in some of the atmospheres at different wavelengths, including \ce{H2O} and \ce{O3}. Oxygen-dominated atmospheres generate ozone photochemically, and ozone has a 9.6~\um{} band that may be present, but was {generally} only visible in the thicker ($\ge$1~bar) atmospheric models. {Except the pure \ce{O2} case,} the 0.1~bar models did not generate sufficient ozone to exhibit a distinctive absorption feature. The 6~\um{} water vapor band is strong, and was present in all of the \ce{water-containing} atmospheres modeled in this work. This band had sufficient absorption to result in nearly no secondary eclipse depth, and would be unlikely to be useful to distinguish among the \ce{O2}-dominated atmospheres considered here.

\section{Discussion}

\begin{figure*}
\centering
\includegraphics[width=\textwidth]{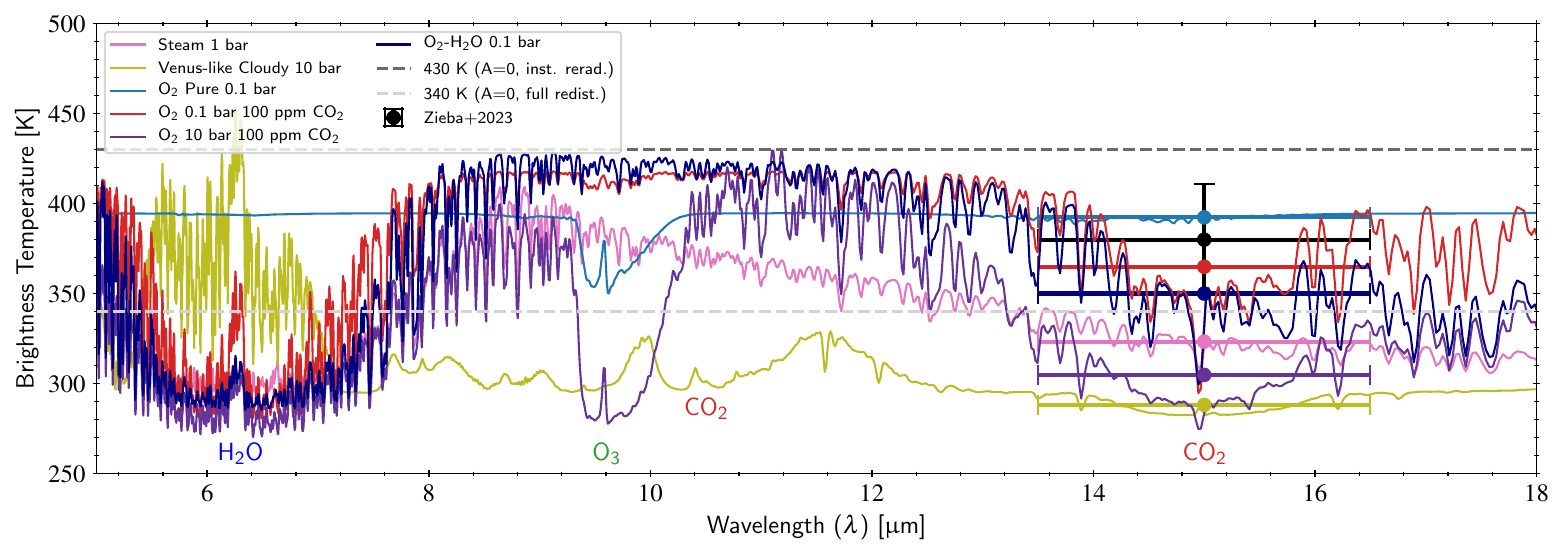}
\caption{Brightness temperature spectra for the dayside hemisphere of  selected modeled environments for TRAPPIST-1~c with \textit{JWST} with points corresponding to the model spectra convolved to the F1500W filter band, over the full-wide half-max of the band's wavelength extent (horizontal error bars). We also show the data point measured by \citet{Zieba:2023}, along with lines corresponding to blackbody temperatures for day-side with zero heat redistribution (dark grey dashed) and for a blackbody with perfect heat redistribution (light gray dashed). The best fit atmospheres are the 0.1~bar, \ce{O2}-dominated atmospheres, at 0.5--1.0$\sigma$ from the measured eclipse depth.}
\label{fig:eclipse_selected}
\end{figure*}

In this work, we modeled a selection of 0.1--10~bar atmospheres dominated by \ce{CO2}, \ce{H2O}, and \ce{O2}, and found that several of them produce 15~\um{} secondary eclipse depths that are within 0.5--2$\sigma$ of the observation \citep{Zieba:2023}, and therefore the possible presence of an atmosphere on TRAPPIST-1 c cannot currently be ruled out. The most interesting and archetypal of our modeled atmospheres are plotted together in Figure~\ref{fig:eclipse_selected}. {Three} of our modeled environments, the {pure 0.1~bar \ce{O2} atmosphere, the} 0.1~bar \ce{O2}/100~ppm \ce{CO2} atmosphere, and the 89\% \ce{O2}/10\% \ce{H2O} atmosphere, are within the 1$\sigma$ secondary eclipse measurement error.  Our 0.1 to 3 bar steam atmospheres were all within {1.7--1.8$\sigma$} of the secondary eclipse value, and so cannot be conclusively ruled out. More massive 1--10~bar \ce{O2} atmospheres, with 100~ppm of \ce{CO2}, are within 2$\sigma$ of the secondary eclipse value. 

For the purposes of understanding the {secondary eclipse depths in the 15~\um{} band for terrestrial atmospheres}, where \ce{CO2} exhibits a substantial absorption band, our modeled atmospheres can be split into two groups: either the atmosphere is dominated by a key radiative gas (\ce{H2O}, \ce{CO2}), or dominated by a transparent gas (\ce{O2}) with radiatively-active trace gases. {Generally, advection transports heat from the day side to the night side, cooling the dayside atmosphere and potentially decreasing the secondary eclipse depth. However, when that observation is centered on a molecular absorption band, as it is here, radiative processes can potentially play a more dominant role. Higher pressure atmospheres can more efficiently redistribute heat, leading to cooler dayside and higher nightside temperatures \citep{Koll:2022}. We clearly see this mechanism operating in our simulations, via the higher nightside outgoing longwave radiation (OLR; Table \ref{table:experiments}) for higher pressure atmospheres.  However, due to radiative effects, we are not sensitive to this behavior in the 15~\um{} secondary eclipse depths. For example, the Venus-like atmospheres show increasing nightside OLR indicating more efficient heat transport with increasing atmospheric pressure. This effect is also seen for the steam atmospheres, although it is less pronounced, due to the fundamental limits on OLR induced by the steam-mediated greenhouse effect \citep{Nakajima:1992}.  Furthermore, these atmospheres are radiatively-dominated by \ce{CO2} and \ce{H2O} opacity, which fixes the emitting altitude at a common or similar lower pressure level, higher in the atmospheres. The atmosphere below this level, where the majority of heat transport occurs, has no observable effect on the emission spectra.  The eclipse depth is then dominated by the atmospheric gases that are spectrally-active in the 15~\um{}, JWST F1500W filter bandpass, and the day-side temperature structure in the radiative region of the atmosphere. Due to this effect, the clear-sky Venus atmospheres exhibited very similar \ce{CO2} eclipse depths, and even a thin 0.1~bar steam atmosphere exhibits a near-identical MIR emission spectrum to the 1--3~bar steam atmospheres.  In comparison, the optically thin \ce{O2}-dominated atmospheres show the perhaps expected effect of higher pressure atmospheres producing more efficient transport, coupled with smaller secondary eclipse depths.  This trend is enabled by lower \ce{CO2} optical depths, which allow emission from different levels of the atmosphere.} 

Our results also illustrate the importance of self-consistent calculation of photochemistry, atmospheric composition, and the vertical temperature profile when attempting to predict secondary eclipse depths. The atmospheric vertical structure can strongly affect the strength of molecular absorption in emission, and this was shown in our calculations of both the Venus-like and \ce{O2}-\ce{CO2} atmospheres.  Although we might have expected the clear-sky Venus atmospheres to have allowed access to hotter, lower levels of the atmosphere, the comparison between our modeled clear and cloudy Venuses showed that the cloudy Venus in fact had a larger eclipse depth, due to the effect of localized heating by clouds in the 15~\um{}-emitting region of the atmosphere. While the cloudy Venus-like atmosphere was 2.6$\sigma$ from the observed value, the clear-sky cases were 3.0--3.1$\sigma$.

Our comparison of the \citet{Zieba:2023} pure \ce{O2}-\ce{CO2} atmospheres with our photochemically self-consistent, outgassing (including water vapor) counterparts, provide consistent results at lower atmospheric pressures, but deviate at higher pressures.  Results for our best fit model {containing both \ce{O2} and \ce{CO2} }(0.1 bar \ce{O2} with 100~ppm \ce{CO2}) are comparable (0.5$\sigma$ here vs -0.6$\sigma$ in \citealt{Zieba:2023}). However, \citet{Zieba:2023} also ruled out a grid of substantial ($\ge1$~bar) \ce{O2} atmospheres with \ce{CO2} levels $\ge100$~ppm, at up to 4.5$\sigma$. Yet with hotter temperature structures derived from full radiative-convective physics and our more complex atmospheric compositions---which include the climatic effect of outgassed water---our results are less pessimistic: we found that the 1--10~bar \ce{O2}, 100~ppm \ce{CO2} are 2.0--2.2$\sigma$ from the measurement, with our most pessimistic \ce{O2} case at 2.9$\sigma$. Additionally, several of the \citet{Zieba:2023} atmospheric models exhibited secondary eclipse depths higher than the measurement, while none of our models exceeded the measurement.

Our modeling also indicated that a steam atmosphere for TRAPPIST-1 c, at {1.7--1.8$\sigma$}, could not be conclusively ruled out, and we derived a rough limit on best fit water abundance for optically thin atmospheres.  Although atmospheric escape rates are likely high for TRAPPIST-1 c \citep{Wordsworth:2018,Dong:2018}, ongoing outgassing from a volatile rich interior may help to maintain an atmosphere against loss processes, including ion loss, as has been argued for other M dwarf planets \citep{Garcia:2017,Wordsworth:2018}. We found that, when compared to atmospheres with different bulk composition, the steam atmospheres can suppress the detectability of \ce{CO2}. For example, the 0.1 bar steam atmosphere with relatively high levels of \ce{CO2} (0.1\%) does not show a \ce{CO2} absorption band at 15~\um{}.  Conversely, the 0.1~bar \ce{O2} atmosphere with 100~ppm \ce{CO2}---which is our best fit (0.5-$\sigma$) candidate for the environments that we considered---can have additional water, up to 10\%, and still be within 1$\sigma$ of the measured eclipse depth.  This atmosphere therefore represents a rough upper limit for atmospheric water vapor inventory for a 1$\sigma$ fit to the data, assuming an optically thin TRAPPIST-1~c atmosphere. Additional water vapor increases the opacity and reduces the eclipse depth, up until reaching a saturation point between 0.01 and 0.1~bar of water, where further water vapor no longer affects the 15~\um{} eclipse depth---as demonstrated by the $\ge$ 0.1~bar steam atmospheres.  

In the future, to better determine the presence and nature of an atmosphere on TRAPPIST-1~c, it will be critical to improve the precision of the 15~\um{} measurement as well as obtaining eclipse depths at other wavelengths to help distinguish between rocky surfaces and different types of atmospheres. Thicker atmospheres are more likely to have other MIR absorption bands (e.g.~\ce{CO2}, \ce{O3}), but their expected deeper 15~\um{} secondary eclipse depths make them less likely, considering the current measurement. MIR observations at 6~\um{}, if feasible, could indicate the presence of water vapor, even at relatively low abundance. Additional measurements between 8--12~\um{} (particularly JWST/MIRI F1000W), whether establishing some molecular absorption or a radiatively clear state, may also help distinguish between different atmospheric states and a rocky surface. NIR transit spectroscopy should also be considered, as it may be better suited to detect atmospheric absorption than further MIR observations \citep[see e.g.][]{Lincowski:2018,Lustig:2019}.

\section{Conclusions}

We simulated a selected variety of plausible post- or ongoing-water loss atmospheres for TRAPPIST-1~c, and compared their secondary eclipse spectra to the measured secondary eclipse depth of \citet{Zieba:2023} to assess compatibility with the data. We broadly considered Venus-like, steam/water-rich, and oxygen-dominated atmospheres.  {Confirming the results of \citet{Zieba:2023}, we} find that the data do not conclusively rule out thin, radiatively transparent atmospheres, such as 0.1~bar \ce{O2}-dominated environments, with low \ce{CO2} abundance which fall within the 1$\sigma$ error bars of the 15~\um{} secondary eclipse measurement of \citet{Zieba:2023}. {We also find that} a maximum of approximately 10\% \ce{H2O} is {consistent with the data to within 1$\sigma$}. Similarly, we find that steam atmospheres of $\geq$0.1~bar are within {1.7--1.8$\sigma$, and are not ruled out by the observation}. Thick \ce{O2} atmospheres are also possible, but less likely at 2.2--2.9$\sigma$ {although these results are less pessimistic than those for similar atmospheres in \citet{Zieba:2023}.} Venus-like atmospheres of $\geq$0.1~bar are excluded at 2.6--3.1$\sigma$ confidence. {We also show that both molecular radiative and heat transport effects need to be considered when estimating or interpreting secondary eclipse depths, and that optically-thick atmospheres can reduce sensitivity to day-night heat redistribution.} Future observations of TRAPPIST-1~c should include other spectral ranges or methods, such as NIR transit spectroscopy, additional MIRI observations in other filter bands, or observation of a MIR phase curve. These observations would help capture other molecular bands, and constrain the rate at which the planet cools. These future observations may help differentiate between a bare rocky surface and an atmosphere, and potentially further constrain the composition of a potential atmosphere. 

\section{Acknowledgements}

This work is based in part on observations made with the NASA/ESA/CSA James Webb Space Telescope. The data were obtained from the Mikulski Archive for Space Telescopes at the Space Telescope Science Institute, which is operated by the Association of Universities for Research in Astronomy, Inc., under NASA contract NAS 5-03127 for JWST. These observations are associated with program \#2304.  A.L. and V.M. are supported by the Virtual Planetary Laboratory Team, which is a member of the NASA Nexus for Exoplanet System Science, and funded via NASA Astrobiology Program Grant 80NSSC18K0829. M.G. is F.R.S.-FNRS Research Director; M.G. and E.D. acknowledge support from the Belgian Federal Science Policy Office BELSPO BRAIN 2.0 (Belgian Research Action through Interdisciplinary Networks) for the project PORTAL n° B2/212/P1/PORTAL789 (PhOtotrophy on Rocky habiTAble pLanets). E.D. also acknowledges support from the innovation and research Horizon 2020 program in the context of the  Marie Sklodowska-Curie subvention 945298. This work made use of the advanced computational, storage, and networking infrastructure provided by the Hyak supercomputer system at the University of Washington. We thank Eric Wolf for providing 3D GCM results from ExoCAM. We thank the anonymous reviewer for thorough and helpful comments, which have improved this work.

\software{Matplotlib \citep{Hunter:2007}, Numpy \citep{Harris:2020}, LBLABC \citep{Meadows:1996}, DISORT \citep{Stamnes:1988,Stamnes:2000}, SMART \citep{Meadows:1996}, VPL Climate \citep{Meadows:2018a,Robinson:2018,Lincowski:2018}, GNU Parallel \citep{Tange:2011}}

\bibliographystyle{aasjournal} \bibliography{linc}

\appendix

\section{VPL 2-column (1.5D) Climate Model} \label{app:twocol}

The two-column mode for our 1D climate model, VPL Climate, was first introduced in \citet{Lincowski:PhD}. A full modeling paper is in preparation at the time of publication of this paper. Here we provide necessary description on our day-night model. 

\subsection{Advection Calculation}

To calculate day-night heat transport, we simplify the quasi-static forms of the primitive equations of atmospheric motion for use in a two-column framework \citep[see][]{Lincowski:PhD}, and here we focus on the horizontal winds:
\begin{equation} \label{eq:horizontal_wind}
    \frac{\partial u}{\partial t} = u \frac{\partial w}{\partial z} - w \frac{\partial u}{\partial z} - \frac{\partial \Phi}{\partial \lambda} - uF_D,
\end{equation}
where $u$ is the horizontal wind, the layer-by-layer variable we solve for in this equation (all variables here are a function of $z$, the altitude), and $\lambda$ is the horizontal distance scale, which here we take to be $\pi r$, where $r$ is the planetary radius. The vertical wind $w$ is derived from our mixing length convection code, from the night side. Here, $\frac{\partial \Phi}{\partial x}$ is the difference in geopotential between the day and night hemispheres for each layer. The primitive equations are not necessarily meant for use for a hemispheric-scale grid, which we find especially problematic for the geopotential. Therefore, we specify a decay component as follows:
\begin{equation}
    \Delta\Phi= (\Phi_{night}-\Phi_{day}) (1-\exp(-H/z)),
\end{equation}
where $H$ is the scale height. Lastly, a frictional term is required, and we use a form of Rayleigh friction, directly proportional to the horizontal wind velocity:
\begin{equation}
    F_D =  \left(D_0 \frac{\rho}{P_\text{orb}} + D_1\right),
\end{equation}
where $D_0 = 1$~kg~m$^{-3}$, $D_1 = 10^{-6}$~s$^{-1}$, and $P_\text{orb}$ is the orbital period, in seconds.

We solve equation \ref{eq:horizontal_wind}, assuming steady state, using tridiagonal methods \citep{Press:1996}, which provides a horizontal wind profile. The advective heating can then be calculated by:
\begin{equation}
  q_\text{ad} = u \frac{\partial T}{\partial \lambda},
\end{equation}
where $T$ is the day- or night-side layer temperature. This provides the heating rate due to advection for each layer.

\subsection{Benchmark with 3D GCMs}

As part of the validation for our two-column model, we compared our model results to 3D GCM results. The full behavior will be discussed in a future paper.

Significant GCM work has focused on TRAPPIST-1~e, widely regarded as the most likely to be temperate and potentially habitable in the TRAPPIST-1 system. The TRAPPIST-1 Habitable Atmosphere Intercomparison (THAI) has been published to compare 3D GCM behavior and results for TRAPPIST-1~e. \citet{Fauchez:THAI} laid out the intercomparison and modeling parameters in advance of the modeling work. \citet{Turbet:2022} showed the results for the dry cases, which we compare here with our VPL Climate two-column model, following the same assumptions where possible. Particularly, there were two dry ``benchmark'' cases, ben1 and ben2. Ben1 was modeled as a 1~bar, nitrogen atmosphere with 400~ppm \ce{CO2}. Ben2 was modeled as a 1~bar \ce{CO2} atmosphere. These represent radiatively transparent and opaque atmospheres, respectively.

The THAI results by \citet{Turbet:2022} clearly showed there is some variation between 3D GCMs even with modeling assumptions  made as similar as possible. Considering the three GCMs of similar spatial resolution (ExoCAM, LMD-G, and ROCKE3D): for ben1, there was a spread of 160--164~W~m$^{-2}$ mean OLR, a range of 47--64~W~m$^{-2}$ minimum OLR, and a maximum OLR range of 419--477~W~m$^{-2}$. For the same three models, ben2 had a spread of 174--184~W~m$^{-2}$ mean OLR, 93--129~W~m$^{-2}$ minimum OLR, and maximum OLR of 301--335~W~m$^{-2}$. The minima and maximum depend on spatial resolution, but even so, these represent some significant differences, though they may just represent a single latitude-longitude grid point. 

E. Wolf kindly provided data for the ExoCAM results presented in \citet{Turbet:2022}. To directly compare with our two-column model, which is composed of day and night hemispheres, we integrate the day and night hemispheres for the ExoCAM GCM results. For ben1, the day-side OLR was 238~W~m$^{-2}$ and the night-side was 84~W~m$^{-2}$. This compares to the VPL Climate two-column results of 242~W~m$^{-2}$ and 88~W~m$^{-2}$, respectively. These values are only 4~W~m$^{-2}$ higher than for ExoCAM. Even analyses of Earth's OLR have a few W~m$^{-2}$ differences \citep{Trenberth:2009}. While the ranges for minima and maximum are not strictly comparable with the hemispherical averages, the range for the GCMs was much larger than 4~W~m$^{-2}$.
For ben2, the GCM day-side was 214~W~m$^{-2}$ and the night-side was 146~W~m$^{-2}$. This compares to the VPL Climate two-column results of 225~W~m$^{-2}$ and 150~W~m$^{-2}$, respectively. This difference on the day-side is a bit larger, 11~W~m$^{-2}$, and only 4~W~m$^{-2}$ on the night side. These do provide similar day-night differences in OLR, indicating overall that global transport is substantially similar.

\begin{figure*}
    \centering
    \includegraphics[width=0.45\textwidth]{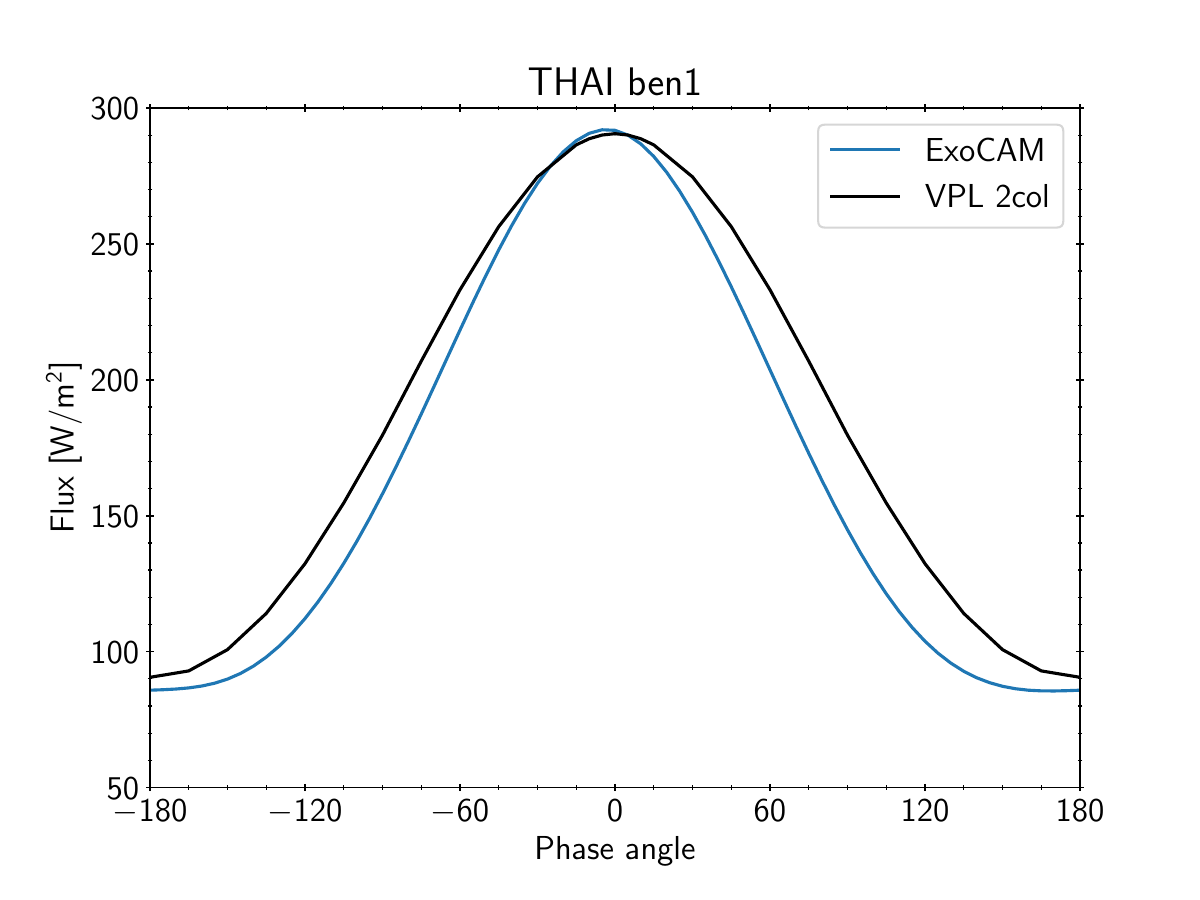}
    \includegraphics[width=0.45\textwidth]{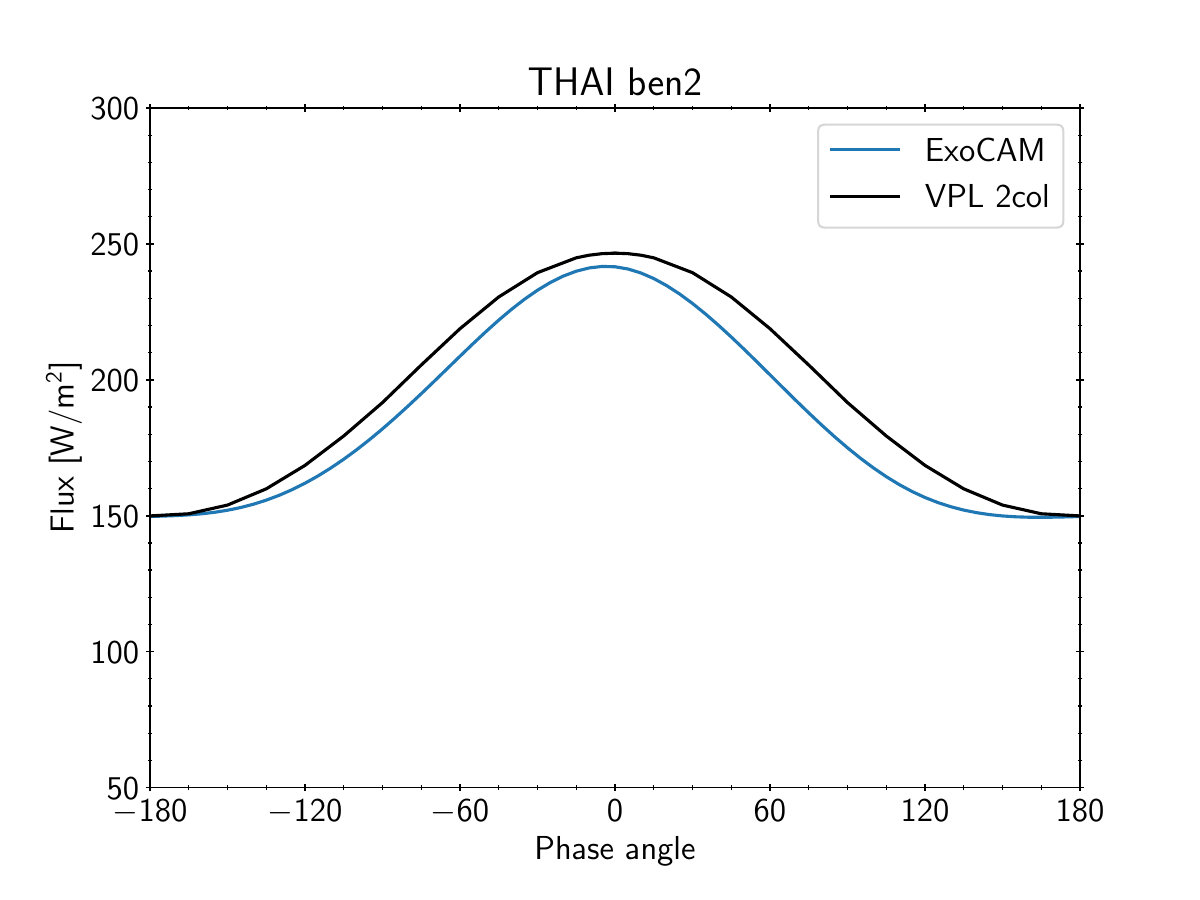}
    \caption{Phasecurves of THAI benchmark cases ben1 and ben2 with VPL Climate (black lines) and ExoCAM (blue lines). These phasecurves demonstrate good agreement between the 3D GCM and two-column models, indicating the day-night advection functions appropriately in VPL Climate.  \label{fig:phasecurves}}
\end{figure*}

In Figure~\ref{fig:phasecurves}, we plot bolometric phasecurves for these ben1 and ben2 test cases, showing the VPL Climate two-column results (black lines) and ExoCAM 3D GCM results (blue lines). There is good agreement between VPL Climate and ExoCAM in these test cases, in both amplitude and total day-side and night-side fluxes. These results indicate the day-night advection in VPL Climate is working appropriately. VPL Climate does produce a slightly warmer phase curve, but the peak and minimum values are similar. The GCM shows a slight offset in the peak flux in each case, which cannot be replicated with a two-column model. 

\section{Photochemistry Results} \label{app:photochem}

\begin{figure*}
    \centering
    \includegraphics[width=0.9\textwidth]{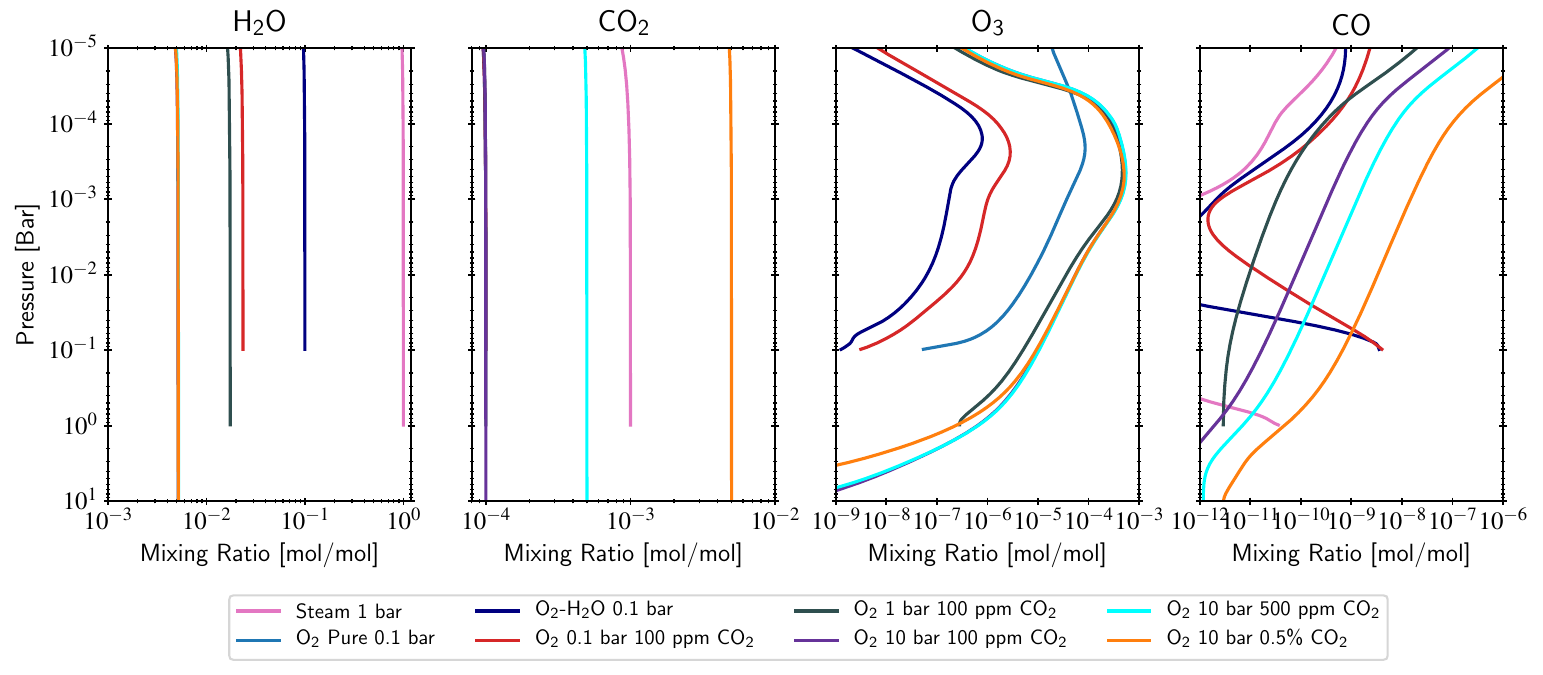}
    \caption{Global mixing ratio profiles for modeled steam (1~bar) and \ce{O2} atmospheres. \ce{H2O} and \ce{CO2} maintain evenly mixed profiles for each case. Ozone and CO are photochemically generated. As discussed in the main text, some of the cases do have CO outgassing, as seen here in the profiles. \label{fig:mixes}}
\end{figure*}

Because the individual trace gases are not very relevant to the 15~\um{} JWST band, which is completely dominated by \ce{CO2} in the majority of our model atmospheres, we have not included the mixing ratio profiles for atmospheric trace gases for our models in the main body of this paper. We present the core gases relevant to the temperature profiles in this Appendix. This information may be relevant for considering other wavelength bands accessible to other JWST filters and instruments, and for future intermodel comparisons. Profiles for \ce{H2O}, \ce{CO2}, \ce{O3}, and CO for the \ce{O2} and steam atmospheres are shown in Figure~\ref{fig:mixes}. Note for the steam atmospheres, we show only the 1 bar profiles. The 0.1 and 3 bar profiles are essentially identical.

The majority of our model atmospheres have water vapor outgassing. Water vapor, in addition to \ce{CO2}, dominate the radiative impact on the temperature profiles. While \ce{H2O} does show up as direct absorption in the 15~\um{} band, it also significantly impacts the temperature structure, which also directly impacts the \ce{CO2} 15~\um{} band absorption. The water profiles in these atmospheres are generally well-mixed, and are shown in the upper left panel of Figure~\ref{fig:mixes}. 

Ozone and CO (Figure~\ref{fig:mixes}, bottom row) are more photochemically generated/mediated in these model atmospheres than \ce{O2}, \ce{CO2}, or \ce{H2O}. Ozone is wholly generated by photochemistry, primarily as a result of photolysis of oxygen, followed by combination of \ce{O2} and O. In water-rich atmospheres, OH serves to react with atomic oxygen also, which limits the generation of ozone. therefore, we see the dry, pure \ce{O2} atmosphere generate more ozone than its other 0.1 bar counterparts. The more water vapor, the less ozone. The 1--10 bar atmospheres all produce similar profiles of ozone. Note here that ozone dry deposition makes a difference. For an Earth-like atmosphere, with abundant near-surface water vapor, we find that chemical loss dominates over dry deposition. Here, we have specified a dry deposition rate for ozone of 0.4~m~s$^{-1}$ \citep{Hauglustaine:1994}, which primarily affects just the pure \ce{O2} case, and only in the lower atmosphere.

The cases with and without CO outgassing can be distinguished by the lower atmosphere profiles. In all cases, CO increases into the upper atmosphere due to \ce{CO2} photolysis. The cases with CO outgassing drop rapidly from the surface before increasing. Unfortunately, CO is generally not observable and does not impact the thermal structure, as the main CO bands are at 2.3 and 4.6~\um{}.

\begin{figure*}
    \centering
    \includegraphics[width=0.9\textwidth]{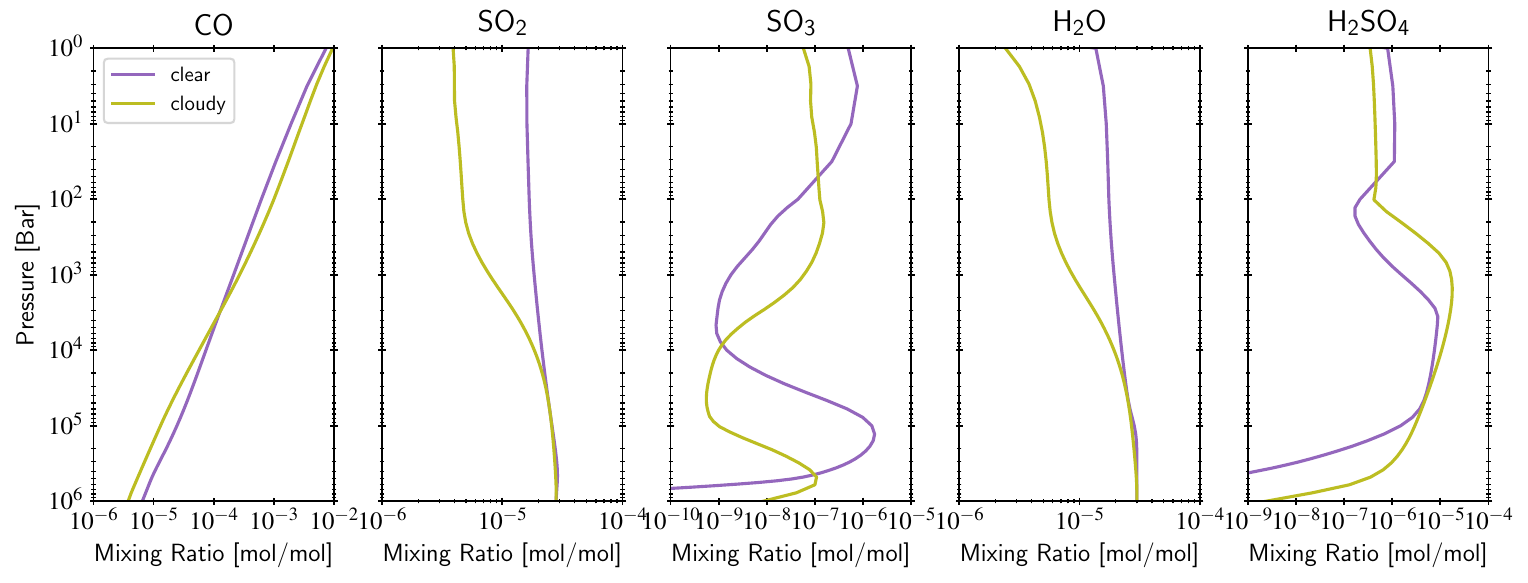}
    \caption{Global mixing ratio profiles for modeled Venus-like atmospheres. Here we show the clear and cloudy 10~bar cases. The lower-pressure clear-sky cases are similar to the 10~bar case. \label{fig:mixes_venus}}
\end{figure*}

In Figure~\ref{fig:mixes_venus}, we plot mixing ratio profiles for key gases related to climate and aerosol formation (CO, \ce{SO2}, \ce{SO3}, \ce{H2O}, and \ce{H2SO4}). Although we use a different stellar SED than in \citet{Lincowski:2018}, these profiles are largely the same, particularly the clear-sky case. One difference observed was in the aerosol-related profiles for the cloudy Venus (\ce{H2O}, \ce{SO2}, and \ce{H2SO4}). This is likely due to increased aerosol formation compared to \citet{Lincowski:2018}. The cloudy profiles for \ce{H2O} and \ce{SO2} are very similar to Venus itself. As was noted in \citet{Lincowski:2018}, TRAPPIST-1~c lies on the cusp of generating \ce{H2SO4} clouds. We continue to find that to be the case here, as the coupled models have difficulty finding a stable state. That is, warming by \ce{H2SO4} aerosols in the climate model evaporates the aerosols, and then leads to cooling, which leads to condensation. Other factors, including uncertainties in the stellar SED and chemical reaction rates, along with other gases or sources of outgassing, can also influence whether \ce{H2SO4} aerosols form. Therefore, we have modeled clear and cloudy states. Future work that more closely couples the climate and photochemical models could allow generation of a intermediate stable state.

\end{document}